\documentstyle{article}
\renewcommand{\thesection}{\arabic{section}}
\renewcommand{\theequation}{\thesection-\arabic{equation}}
\bibliography{plain}
\title{\bf Coherent and Squeezed States in Shape Invariant  Potentials Obtained from
the Master Function}

\vspace{20mm}
\author{M. A. Jafarizadeh$^{a,b,c}$ \thanks{E-mail:jafarzadeh@ark.tabrizu.ac.ir} , A. Rostami  $^{a}$ \thanks{E-mail:rostami@ark.tabrizu.ac.ir}
\\
\\
$^a${\small Department of Theoretical Physics and Astrophysics, Tabriz University, Tabriz 51664, Iran.} \\
$^b${\small Institute for Studies in Theoretical Physics and Mathematics, Tehran 19395-1795, Iran.} \\
$^c${\small Pure and Applied Science Research Center, Tabriz 51664, Iran.}}
\pagebreak

\begin{document}
\maketitle
\vspace{15mm}
\newpage
\begin{abstract}
A general algorithm has been given for the generation of Coherent and Squeezed
states,
in one-dimensional hamiltonians with shape invariant  potential, obtained from
the master function. The minimum uncertainty states of these potentials are
expressed in terms of the well-known special functions.

{\bf Keywords: Squeezed States, Coherent States, Shape Invariance, Special
\hspace{1.2 cm}\hspace{0.9 cm}  Functions. }

{\bf PACs Index: 42.50.Dv, 02.30. +g  or  02.90. +p.}

\end{abstract}
\pagebreak

\vspace{7cm}

\section{Introduction}

Coherent states, known as the closest states to classical ones, play an
important role in many different contexts of the
theoretical and experimental physics, specially quantum
optics\cite{Sl:Klud,Al:Kluy} and
multiparticle dynamics\cite{La:Laug}. Schrodinger first discovered the
coherent states of the harmonic oscillator
potential in $1926$ \cite{Pe:Schr} and much work has been done since then on
their properties and applications\cite{Sl:Klud,Al:Rost,Al:Kluy}.
The coherent states have also been found in systems with the Lie
group symmety \cite{La:Prel,Ph:Phds,Ph:Phdt}.
Recently, coherent states have been found in special
Hamiltonians \cite{Do:Niet,Do:Niea,Do:Nieb}.
These coherent states are called
minimum uncertainty coherent states(MUCS). In coherent states the
standard
deviation of $x$ and $p$ are equal and their product is minimum over
these states. There are also quantum states where, though we have minimum
uncertainty for the standard deviation of coordinate and momentum, they are
not
equal any more; these states are called squeezed states. These quantum states
are as important as coherent ones and their generation play
an important role in many different branch of physics and communication
engineering\cite{La:Josc,Sl:Sluy}. Nieto and his colleagues have developed an
interesting algorithm
for the generation of the coherent and squeezed states for special potential,
where the product
of the standard deviation generalized Harmonic phase variable $X_c$ and $P_c$
are minimum over these states. Here in this article following the algorithm
of reference \cite{Do:Niet,Do:Niea,Do:Nieb} we obtain these coherent and squeezed states for
all the shape
invariant potential obtained from the master function of reference
\cite{Lo:Jafa}.
The Hamiltonians of the reference\cite{Do:Niea,Do:Nieb} are the special cases of
the most
general shape invariant Hamiltonian to be treated in this paper and
the result thus obtained in this work, are in good agreement with those
of thier reference\cite{Do:Niea,Do:Nieb} in these few special cases.

This paper is organized as follows.
In section II we explain very briefly the shape invariant potential
obtained from the master functions. In section III first we show that the
generalized Harmonic variable $X_c$  of reference \cite{Do:Niet} is
 linear function of the $x$ coordinate of orthogonal function and
$P_c$ the generalized  Harmonic momentum is proportional to master function.
In section IV following the reference \cite{Do:Niet}, we obtain the raising
and lowering operator of these potentials. In section V we obtain the most
general minimum uncertainty states for the general potential obtained from
the master function. For particular choice of even master and weight function
together with the symmetric interval we can obtain even or odd minimum
uncertainty coherent (squeezed)states or well known Schrodinger-Cat state.
It is also shown that the ground state of the Hamiltonian is one of
the MUCS of the system.
Eigenstates of Annihilation operators namlely the generalization of
annihilation
operators coherent states(AOCS) is derived in section VI. Here in this
section
it is shown that in general the minimum uncertainty states  are different
from AOCS ones.
Section VII devoted to investigate the time evolution of the Minimum
Uncertainty states. Here in this section it is shown that, the time evolution
of the generlized quantum phase coordinates is almost similar to the time
evolution of the phase coordinates of the qunantum oscillator except for the
appearance of the constant phase ${\omega_0}$ and also the Hamiltonian
depenence of the frequency $\omega_H$. The paper ended with a brief conclusion.

\begin{center}
\section{The Shape InvariantPotentials Obtained From the Master function}
\end{center}

According to references \cite{Lu:Jafa} by introducing the master function
$A(x)$ as a
polynomial of at most second order one can define a non-negative weight
function $W(x)$ in the interval $[a,b]$ such that the expression
$\frac{1}{W(x)}\frac{d}{dx}(A(x)W(x))$ be a polynomial of at most
first order
and the function $A(x)W(x)$ to vanish at the ends of the interval.
Now we can define second order differential operator
$L=\frac{1}{W(x)}\frac{d}{dx}A(x)W(x)\frac{d}{dx}$ with the following
properties:\\
1) $L$ is a self-adjoint linear operator.\\
2) $L$ transforms a given polynomial of order $m$ to another polynomial
   of order $m$ at most.\\
3) The expression $\frac{1}{W(x)}(\frac{d}{dx})^{^n}(A^{^n}(x)W(x))$ is a
   polynomial of order at most $n$,
   which is indeed Rodrigues formula for the classical orthogonal polynomials.\\
4) The polynomials
$$\phi_n(x)=\frac{a_n}{W(x)}(\frac{d}{dx})^{^n}(A^{^n}(x)W(x))$$
   are orthogonal with respect to the weight function $W(x)$ in the interval $[a,b]$ as
   defined above, and one can find $a_n$ simply by comparing the coefficient
   of highest power of $\phi_n(x)$ with those of the traditionally defined
   special orthogonal polynomials.\\
5) The polynomials $\phi_n(x)$ are eigenfunctions of operator $L$, and therefore satisfy the following second order
   linear differential equation

\begin{equation}
\frac{1}{W(x)}\frac{d}{dx}(A(x)W(x)\frac{d}{dx}\phi_n(x))=-\gamma_n\phi_n(x).
\end{equation}

In order the differential $Eq.(2-1)$ have polynomial solution of degree $n$,
$\gamma_n$ must be given by

$$
\gamma_n=-n(\frac{(A(x)W(x))^{\prime}}{W(x)})^{\prime}-\frac{n(n-1)}{2}A^{\prime\prime}(x).
$$

Thus, the general form of the differential equation is as follows

\begin{equation}
A(x)\phi_n^{\prime\prime}(x)+\frac{(A(x)W(x))^{\prime}}{W(x)}\phi_n^{\prime}(x)
-[n(\frac{(A(x)W(x))^{\prime}}{W(x)})^{\prime}+\frac{n(n-1)}{2}A^{\prime\prime}(x)]
\phi_n(x)=0.
\end{equation}

By differentiating the differential Eq.$(2-2)$ $m$ times and then multiplying
it by $(-1)^{^m}A^{\frac{m}{2}}(x)$ we get the following associated differential
equation
$$
A(x)\phi^{\prime\prime}_{_{n,m}}(x)+\frac{(A(x)W(x))^{\prime}}{W(x)}\phi^{\prime}_{_{n,m}}(x)+[-\frac{1}{2}(n^2+n-m^2)A^{\prime\prime}(x)
+(m-n)(\frac{A(x)W^{\prime}(x)}{W(x)})^{\prime}
$$
\begin{equation}
-\frac{m^2}{4}\frac{(A^{\prime}(x))^{^2}}{A(x)}-\frac{m}{2}\frac{A^{\prime}(x)W^{\prime}(x)}{W(x)}]\phi_{_{n,m}}(x)=0
\end{equation}
where
$$
\phi_{_{n,m}}(x)=(-1)^{^m} A^{\frac{m}{2}}(x)(\frac{d}{dx})^{^m}\phi_{_n}(x).
$$
Now, changing the variable $\frac{dx}{d\xi}=\sqrt{A(x)}$ in associated differential
equations of Eq.$(2-3)$ and defining the new function
$\psi^{m}_n(\xi)=A^{\frac{1}{4}}(x)W^{\frac{1}{2}}(x)\phi_{n,m}(x)$
we obtain the Schr\"{o}dinger
equation \cite{Lo:Jafa}

\begin{equation}
-\frac{d^2}{d\xi^2}\psi^{m}_n(\xi)+V_m(x(\xi))\psi^{m}_n(\xi)=E(n,m)\psi^{m}_n(\xi),  \hspace{10mm}  m=0,1,2, \cdots ,n ,
\end{equation}
where the most general shape invariant potential is

$$
V_m(x)=W^2(x)+\frac{d}{d\xi}W(x)=-\frac{1}{2}(\frac{A(x)W^{\prime}(x)}{W(x)})^{\prime}-\frac{2m-1}{4}A^{\prime\prime}(x)+\frac{1}{4A(x)}(\frac{A(x)W^{\prime}(x)}{W(x)})^2+
$$
\begin{equation}
\frac{m}{2}\frac{A^{\prime}(x)W^{\prime}(x)}{W(x)}
+\frac{4m^2-1}{16}\frac{A^{\prime 2}(x)}{A(x)},
\end{equation}
and prime stands for derivative with respect to $x$. The spectrum $E(n,m)$ is
\begin{equation}
E(n,m)=-(n-m+1)[(\frac{A(x)W^{\prime}(x)}{W(x)})^{\prime}+\frac{1}{2}(n+m)A^{\prime\prime}(x)].
\end{equation}

\section{Generalized Harmonic Phase space variables $X_c$ and $P_c$ }

Following the prescription of references \cite{Do:Niet,Do:Niea,Do:Nieb}
 in a one dimensional hamiltonian:

\setcounter{equation}{0}
\begin{equation}
H=\frac{1}{2}(\frac{d\xi}{dt})^2+V_m(x(\xi)),
\end{equation}
with $V_m(x(\xi))$ given in Eq.$(2-5)$, the classical paths of constant
energy around the minimum points of the potential form closed paths in the
phase space $(\xi,P_{\xi})$. Therefore, there is an injective canonical  map
from the phase space $(\xi,P_{\xi})$ into the new phase space $(X_c,P_c)$ ,
such that the closed constant energy paths turn into elliptic constant energy
paths. Hence, in the phase space  $\xi-P_{\xi}$  the time dependence of
these closed paths can be written as

\begin{equation}
X_c=A(E)\sin \omega_c(E)t,
\end{equation}

\[P_c=mA(E)\omega_c(E)\cos \omega_c(E)t.\]
From the constancy of the hamiltonian, Eq.$(3-1)$, along the paths we have

\[t+t_0=\int \frac{d\xi}{\sqrt{\frac{2}{m}(E-V_m(x(\xi))}}.\]
By changing the variable $\frac{dx}{\sqrt{A(x)}}=d\xi$, we get
\begin{equation}
t+t_0=\sqrt{\frac{m}{2}}\int \frac{dx}{\sqrt{A(x)(E-V_m(x))}}.
\end{equation}
Inserting the expression $(2-5)$ for $V_m(x)$ and considering the fact that $A(x)$
is at most second order and $A(x)\frac{d\log W(x)}{dx}$ is at most
first order, one can show that the expression $A(x)(E-V_m(x))$ is quadratic.
Hence, integrating Eq.$(3-5)$ we obtain
\begin{equation}
x+\frac{\eta_2}{2\eta_1}=\sqrt{(\frac{\eta_2}{2\eta_1})^2-\frac{\eta_3}{\eta_1}}
\sin \sqrt{\frac{-2\eta_1}{m}}(t+t_0)
\end{equation}
with
\[ \eta_1=\frac{1}{2}A^{\prime\prime}(E-\gamma +\frac{2m-1}{4}A^{\prime\prime})
+\frac{1-2m}{4}A^{\prime\prime}(\frac{AW^{\prime}}{W})^{\prime}
-\frac{1}{4}(\frac{AW^{\prime}}{W})^{\prime}
-\frac{4m^2-1}{16}(\frac{AW^{\prime}}{W})^{\prime} \]

\[ \eta_2=A^{\prime}(0)(E-\gamma +\frac{2m-1}{4}A^{\prime\prime})
+\frac{1}{2}A^{\prime}(0)(\frac{AW^{\prime}}{W})^{\prime}
-\frac{1}{2}(\frac{AW^{\prime}}{W})^{\prime}(\frac{AW^{\prime}}{W})(0)
-\frac{m}{2}(A^{\prime\prime}(0)+A^{\prime}(0))\frac{AW^{\prime}}{W})(0)\]

\begin{equation}
\eta_3=A(0)(E-\gamma +\frac{2m-1}{4}A^{\prime\prime})
+\frac{1}{2}A(0)(\frac{AW^{\prime}}{W})^{\prime}
-\frac{1}{4}(\frac{AW^{\prime}}{W}(0))^2
-\frac{m}{2}A^{\prime}(0)(\frac{AW^{\prime}}{W})(0)
\frac{4m^2-1}{16}(A^{\prime}(0))^2.
\end{equation}
Comparing the relations $(3-2)$ and $(3-6)$ we have

\[X_c=x_0(x+\frac{\eta_2}{2\eta_1})\]

\[ \omega_c=\sqrt{\frac{-2\eta_1}{m}}\]

\begin{equation}
A(E)=x_0\sqrt{(\frac{\eta_2}{2\eta_1})^2-\frac{\eta_3}{\eta_1}},
\end{equation}
where $x_0$ and $t_0$ are arbitrary constants of integration. Also $\gamma$
is a constant which is added to the potential $V_m(x(\xi))$ for convenience.
Similarly, for the momentum $P_c=m\frac{dX_c}{dt}$ we have
\begin{equation}
P_c=mx_0\frac{dx}{dt}=x_0m\frac{d\xi}{dt}\frac{dx}{d\xi}=x_0mP_{\xi}\sqrt{A(x)}.
\end{equation}

The quantum operators corresponding to $X_c$ and $P_c$, denoted by
$\widehat{X}$ and $\widehat{P}$, are defined as following according to \cite{Do:Niet,Do:Niea,Do:Nieb}:

\[\widehat{X}=X_c(x)=x_0(x+\frac{\eta_2}{2\eta_1})\]

\[\widehat{P}=\frac{1}{2i}(X_c^{\prime}p+pX_c^{\prime}).\]

Making a change variable from $\xi$ to $x$, we get

\[\widehat{P}=\frac{p_0}{2i}(A(x)\frac{d}{dx}+\frac{d}{dx}A(x))\]
where $x_0$ and $p_0$ are arbitrary constants.

\begin{center}
\section{Raising And Lowering Operators}
\end{center}

In the algorithm of generation of the minimum uncertainty states of the
hamiltonian $-\frac{d^2}{d \xi^2} + V_m(x(\xi))$, the raising and lowering
operators of its discrete eigenstates $\psi^{m}_n$, play an important
role. These operators are denoted by $ \tilde{B}_{n,m}$ and
$ \tilde{A}_{n,m}$ respectively. To obtain them, following the
refernce \cite{Lu:Jafa}, first we factorize the differential equation $(2-3)$
in a shape invariant form as:

\setcounter{equation}{0}
\begin{equation}
\left\{\begin{array}{cc}
B(n,m)A(n,m) \phi_{_{n,m}}(x)=E(n,m) \phi_{_{n,m}}(x) \\
 A(n,m)B(n,m) \phi_{_{n-1,m}}(x)=E(n,m) \phi_{_{n-1,m}}(x).
\end{array}
\right.
\end{equation}

From the equations $(4-1)$ it is straightforward to derive the following recursion
relations

\begin{equation}
\left\{\begin{array}{cc}
B(n,m)\phi_{_{n-1,m}}(x)=\mu_{m,n}\phi_{_{n,m}}(x)\\
A(n,m) \phi_{_{n,m}}(x)= \frac{E(n,m)}{\mu_{m,n}} \phi_{_{n-1,m}}(x),
\end{array}
\right.
\end{equation}

where $E(n,m)$, $B(n,m)$ and $A(n,m)$  are given by

\[E(n,m)=\]
\[\frac{[(\frac{A(x)W^{\prime}(x)}{W(x)})^{\prime}\frac{AW^{\prime}}{W}(0)+n^{ ^2}A^{\prime\prime}(x)A^{\prime}(0)+(2n-m)(\frac{A(x)W^{\prime}(x)}{W(x)})^{\prime}A^{\prime}(0)+mA^{\prime\prime}(x)\frac{AW^{\prime}}{W}(0)]^{^2}}
{4[(\frac{A(x)W^{\prime}(x)}{W(x)})^{\prime}+nA^{\prime\prime}(x)]^{^2}} \]

\[-\frac{1}{4}(\frac{AW^{\prime}}{W}(0))^{^2}-(n-m)(\frac{A(x)W^{\prime}(x)}{W(x)})^{\prime}A(0)-\frac{1}{4}m^{^2}(A^{\prime}(0))^{^2}-\frac{1}{2}mA^{\prime}(0)(\frac{AW^{\prime}}{W})(0)\]

\hspace{6mm}\[-\frac{1}{2}(n^{^2}-m^{^2})A^{\prime\prime}(x)A(0)\]

$$
B(n,m)=-A(x)\frac{d}{dx}-((\frac{A(x)W^{\prime}(x)}{W(x)})^{\prime}+\frac{1}{2}nA^{\prime\prime}(x))x
$$
$$
-\frac{2(\frac{A(x)W^{\prime}(x)}{W(x)})^{\prime}\frac{AW^{\prime}}{W}(0)+n^{^2}A^{\prime\prime}(x)A^{\prime}(0)+(2n-m)(\frac{A(x)W^{\prime}(x)}{W(x)})^{\prime}A^{\prime}(0)+(m+n)A^{\prime\prime}(x)\frac{AW^{\prime}}{W}(0)}{2[
(\frac{A(x)W^{\prime}(x)}{W(x)})^{\prime}+nA^{\prime\prime}(x)]}
$$
\\

$A(n,m)=A(x)\frac{d}{dx}-\frac{1}{2}nA^{\prime\prime}(x)x$
\begin{equation}
-\frac{n^{^2}A^{\prime\prime}(x)A^{\prime}(0)+(2n-m)(\frac{A(x)W^{\prime}(x)}{W(x)})^{\prime}A^{\prime}(0)+(m-n)A^{\prime\prime}(x)\frac{AW^{\prime}}{W}(0)}{2[(\frac{A(x)W^{\prime}(x)}{W(x)})^{\prime}+nA^{\prime\prime}(x)]}.
\end{equation}

In order to evaluate $\mu_{n,m}$, it is sufficient to divide both sides of
Eq.$(4-2)$ by $(A(x))^{\frac{m}{2}}$ and compare the coefficients of
the highest degree terms of boths sides, resulting in

\begin{equation}
\mu_{n,m}=[-\frac{1}{2}A^{\prime\prime}(x)(n-1) - ((\frac{AW^\prime}{W})^{\prime} + \frac{1}{2}nA^{\prime\prime})].
\end{equation}

From $\psi_n^{m}(\xi)=A^{\frac{1}{4}}W^{\frac{1}{2}}\phi_{n,m}$, it follows that
$\widehat{A}_{n,m}$ and $\widehat{B}_{n,m}$, that is

\begin{equation}
\left\{\begin{array}{cc}
\tilde{A}_{n,m}=A^{\frac{1}{4}}W^{\frac{1}{2}} A_{n,m} A^{\frac{-1}{4}}W^{\frac{-1}{2}}\\
 \tilde{B}_{n,m}=A^{\frac{1}{4}}W^{\frac{1}{2}} B_{n,m} A^{\frac{-1}{4}}W^{\frac{-1}{2}}
\end{array}
\right.
\end{equation}

are the required raising and lowering operators of the wave functions,
that is, we have

\begin{equation}
\left\{\begin{array}{cc}
\tilde{B}_{n,m}\psi_{n-1}^{m}(\xi)=\mu_{n,m}\psi_{n}^{m}(\xi)\\
 \tilde{A}_{n,m}\psi_{n}^{m}(\xi)=\frac{E(n,m)}{\mu_{n,m}}\psi_{n-1}^{m}(\xi).
\end{array}
\right.
\end{equation}

where $\tilde{A}_{n,m}$ and $\tilde{B}_{n,m}$ are

\[\tilde{A}=A(x)\frac{d}{dx}-(\frac{1}{2}nA^{\prime\prime}(x)+\frac{1}{4}A^{\prime\prime}(x)+\frac{1}{2}(\frac{A(x)W^{\prime}(x)}{W(x)})^{\prime})x\]
\[-\frac{1}{2}A^{\prime}(0)-\frac{A(x)W^{\prime}(x)}{W(x)}(0)-\]
\[\frac{n^2A^{\prime\prime}(x)A^{\prime}(0)+(2n-m)A1A^{\prime}(0)+(m-n)A^{\prime\prime}(x)C}{2(A1+nA^{\prime\prime}(x))}\]

\[\tilde{B}=-A(x)\frac{d}{dx}-(\frac{1}{2}nA^{\prime\prime}(x)+(\frac{AW^{\prime}}{W(x)})^{\prime}-(\frac{1}{2}\frac{A(x)W^{\prime}(x)}{W(x)})^{\prime}-\frac{1}{4}A^{\prime\prime}(x))x\]
\[+\frac{1}{2}A^{\prime}(0)+\frac{A(x)W^{\prime}(x)}{W(x)}(0)-\]

\begin{equation}
\frac{2A1C+n^2A^{\prime\prime}(x)A^{\prime}(0)+(2n-m)A1A^{\prime}(0)+(m+n)A^{\prime\prime}(x)C}{2(A1+nA^{\prime\prime}(x))}
\end{equation}

with
$$A1=(\frac{AW^{\prime}}{W(x)})^{\prime}
$$
$$
C=\frac{AW^{\prime}}{W}(0).
$$

The raising and lowering operators of the one dimensional shape invariant
potentials, obtained from the master function, and also all other necessary
information in constructing their minimum uncertainty state are given in
Table I.
The Hermitian quantum operators associated with the generalized Harmonic
quantum phase variables $\widehat{X}$ and $\widehat{P}$ can be expressed in terms of the
operators $\tilde{A}, \tilde{B}$ and their Hermitian conjugates as

\[\widehat{X}=x_0[\tilde{A} + \tilde{A}^{\dagger} + \tilde{B} + \tilde{B}^{\dagger}] \]

\begin{equation}
\widehat{P}=\frac{p_0}{2i}[\tilde{A} + \tilde{B}^{\dagger} - (\tilde{A}^{\dagger} + \tilde{B})],
\end{equation}

where $\tilde{A}^{\dagger}$ and $\tilde{B}^{\dagger}$,
the Hermitian conjugate of the raising and lowering operators
are

\[\tilde{A}^{\dagger}=(\frac{f_1-f_3-A^{\prime\prime}(x)}{f_1+f_3})\tilde{A}+(\frac{2f_1-A^{\prime\prime}(x)}{f_1+f_3})\tilde{B}+ \]
\[\frac{(f_2-A^{\prime}(0))(f_1+f_3)-(f_1-A^{\prime\prime}(x))(f_2+f_4)+\frac{1}{2}(f_2-f_4)(f_1+f_3)-\frac{1}{2}(f_1-f_3)(f_2+f_4)}{f_1+f_3}  \]

\[\tilde{B}^{\dagger}=(\frac{2f_3+A^{\prime\prime}(x)}{f_1+f_3})\tilde{A}+(\frac{f_3-f_1+A^{\prime\prime}(x)}{f_1+f_3})\tilde{B}+ \]
\[\frac{(f_4+A^{\prime}(0))(f_1+f_3)-(f_3+A^{\prime\prime}(x))(f_2+f_4)+\frac{1}{2}(f_1-f_3)(f_2+f_4)-\frac{1}{2}(f_2-f_4)(f_1+f_3)}{f_1+f_3}  \]
with $f_1, f_2, f_3$ and $f_4$:

\[f_1=-\frac{1}{2}A^{\prime\prime}(x)-\frac{1}{2}(\frac{A(x)W^{\prime}(x)}{W(x)})^{\prime}-\frac{1}{4}A^{\prime\prime}(x),\]
\[f_2=-\frac{1}{2}\frac{A(x)W^{\prime}(x)}{W(x)}(0)-\frac{1}{4}A^{\prime}(0)
-\frac{n^2A^{\prime\prime}(x)A^{\prime}(0)+(2n-m)(\frac{A(x)W^{\prime}(x)}{W(x)})^{\prime}A^{\prime}(0)+(m-n)A^{\prime\prime}(x)C}{2((\frac{A(x)W^{\prime}(x)}{W(x)})^{\prime}+nA^{\prime\prime}(x))},\]
\[f_3=-\frac{1}{2}\frac{A(x)W^{\prime}(x)}{W(x)}(0)-\frac{1}{2}A^{\prime\prime}(x)+\frac{1}{4}A^{\prime\prime}(x),\]
and
\[f_4=\frac{1}{2}\frac{A(x)W^{\prime}(x)}{W(x)}(0)+\frac{1}{4}A^{\prime}(0)-\]

\begin{equation}
\frac{n^2A^{\prime\prime}(x)A^{\prime}(0)+(2n-m)(\frac{A(x)W^{\prime}(x)}{W(x)})^{\prime}A^{\prime}(0)+(m+n)A^{\prime\prime}(x)C+2(\frac{A(x)W^{\prime}(x)}{W(x)})^{\prime}C}{2((\frac{A(x)W^{\prime}(x)}{W(x)})^{\prime}+nA^{\prime\prime}(x))}.
\end{equation}

Generally speaking, the number $n$ should not appear anywhere and it
must be replaced by the Hamiltonian. This is done by expressing $n$ in terms of
$E(n,m)$ and replacing $E(n,m)$ by the Hamiltonian. Since the set of
eigenfunctions $\psi_n^m$ are complete and we can expand every function
in our Hilbert space in terms of them, therefore it is sufficient
to consider the effect of the operators on these base. Therefore, in order
not to make things too complicated, we do not bother to
replace $n$ in terms of Hamiltonian, as in reference\cite{Do:Niet}, except for
the operators $\widehat{X}$ and $\widehat{P}$ which are going to be functions of
the Hamiltonian.

\begin{table}
\begin{tabular}{|c|c|c|c|c|} \hline
$A(x)$ and & $W(x)$  and                   & $x=x(t)$              & $\mu_{n,m}$                                                   & $V_m(t)$  and                                                                       \\
Name       & Interval of $x$               &                       &                     & $E(n,m)$                                                                            \\  \hline
$1$        & $e^{-\frac{1}{2}\alpha x^2}$  &                       &                   & $\frac{1}{4}\omega^2(t-\frac{2b}{\omega})^2+\frac{\omega}{2}$                       \\
           & $\alpha>0 $                   &                       &                                                            &                                                                                     \\
           &                               &$x=t-\frac{2b}{\alpha}$& $\frac{n-m}{n} \sqrt{n}$      &                                                                                     \\  \cline{2-2} \cline{5-5}
shifted    &                           &                           &                   & $\alpha(n-m+1)$                                                                     \\
oscillator & $-\infty<x<+\infty$           &                       &                                                            &                                                                                     \\
           &                               &                       &                                                            &                                                                                     \\  \hline
$x$        & $x^{\alpha}e^{-\beta x}$      &                       &                  & $\frac{1}{4}\omega^2t^2+\frac{l(l+1)}{t^2}-(l-\frac{1}{2})\omega$                  \\
           & $\alpha>-1$                   &                       &       &   \\
           & $\beta>0$                     &   $x=\frac{1}{4}t^2$  &$-(n-m) \sqrt{\frac{n+\alpha}{n}}$    &                                                                                     \\  \cline{2-2} \cline{5-5}
three &    &                               &                       & $\beta(n-m+1)$              \\
dimensional& $0<x<+\infty$                 &                       &           &                                                                                     \\
oscillator &                               &                       &                                                            &                                                                                     \\  \hline
$x^2$      &$x^{\alpha}e^{-\frac{\beta}{x}}$&                      &        & $A^2+B^2e^{-2t}-B(2A+1)e^{-t}$                                              \\
           & $\alpha<-2$                   &                       &                                                            &                                                                                        \\
           & $\beta>0$                     & $x=e^t$               &$-\frac{(n+m)(n+\alpha)}{2n+\alpha}\times$   &                                                                                       \\  \cline{2-2} \cline{5-5}
Morse      &                               &                       &$\sqrt{\frac{n+\alpha}{n}}$  &  $-(\alpha+n+m)(n-m+1)$                                     \\
           & $0<x<+\infty$                 &                       &              &                                                                                       \\
           &                               &                       &                                                            &                                                                                       \\ \hline
$1+x^2$    &$(1+x^2)^{\alpha}e^{\beta Arctanx}$&                   &    & $A^2+(B^2-A^2-A) sech^2t+$                                                              \\
           & $\alpha<-1$                   &                       &                                                            & $B(2A+1) tanht secht$                                                                   \\
           & $-\infty<\beta<+\infty$       & $x=sinht$             & $\frac{(n-m)(n+2\alpha)}{2n+2\alpha}\times$   &                                                                                        \\  \cline{2-2} \cline{5-5}
Scarf II   &                               &                       &$\sqrt{\frac{2m-1}{2}}$  & $-(2\alpha+n+m)(n-m+1)$                                                                \\
(hyperbolic)& $-\infty<x<+\infty$           &                       &   &                                                                                        \\
           &                               &                       &                                                            &                                                                                        \\ \hline
$x(1-x)$   & $x^{\alpha}{(1-x)^{\beta}}$   &                       &       & $-A^2+(A^2+B^2+A) sec^2t+$                                                       \\
           & $\alpha, \beta>-1$            &                       &                                                            & $-B(2A+1) tant sect$                                                              \\
           &                               & $x=\frac{1+sint}{2}$  &$-\frac{n(n+\alpha+\beta)}{2n+\alpha+\beta}\times$&\\  \cline{2-2}  \cline{5-5}
Scarf I    &                               &                       &$\sqrt{\frac{(n+\alpha)(n+\beta)(2n+\alpha+\beta-1)}{n(n+\alpha+\beta)(2n+\alpha+\beta+1)}}$&$(\alpha+\beta+n+m)(n-m+1)$   \\
(trigonometric) & $0<x<+1$                 &                       &    &                                                                                   \\
           &                               &                       &                                                            &                                                                                       \\ \hline
$x^2-1$    & $(x-1)^{\alpha}(x+1)^{\beta}$ &                       & & $A^2+(A^2+B^2+A) cosech^2t$                                                       \\
           & $\alpha, \beta>-1$            &                       &$\frac{2n(n+\alpha+\beta)}{2n+\alpha+\beta}\times$ & $-B(2A+1) cotanht cosecht$                                                        \\
           &                               & $x=cosht$             &$\sqrt{\frac{(n+\alpha)(n+\beta)(2n+\alpha+\beta-1)}{n(n+\alpha+\beta)(2n+\alpha+\beta+1)}}$ & \\  \cline{2-2}  \cline{5-5}
generalized&                               &                       & & $-(\alpha+\beta+n+m)(n-m+1)$                                                     \\
P\"{o}schl-teller & $-1<x<+1$                 &                       &                        &                                                                                  \\
           &                               &                       &                               & \\  \hline
$1-x^2$    & $(1-x)^{\alpha}(1+x)^{\beta}$ &                       & & $-A^2+(A^2+B^2-A) cosec^2t$      \\
           & $\alpha, \beta>-1$            &                       &$2\frac{n(n+\alpha+\beta)}{2n+\alpha+\beta}\times$ & $-B(2A-1) cotant cosect$        \\
           &                               &                       & &\\  \cline{2-2} \cline{5-5}
           &                               & $x=cost$              &$\sqrt{\frac{(n+\alpha)(n+\beta)(2n+\alpha+\beta-1)}{n(n+\alpha+\beta)(2n+\alpha+\beta+1)}}$& $(\alpha+\beta+n+m)(n-m+1)$ \\
           & $-1<x<+1$                     &                       &              &                                                                                       \\
           &                               &                       &              &                                                                                       \\   \hline
$4x^2-1$   &$(2x-1)^{\alpha}(2x+1)^{\beta}$&                       &              & $(A-B)^2+\frac{B(B-1)}{sinh^2t}-\frac{A(A+1)}{cosh^2t}$                            \\
           & $\alpha, \beta>-1$            &                       &$\frac{2n(n+\alpha+\beta)}{2n+\alpha+\beta}\times$&                                                                                    \\
           &                               & $x=\frac{1}{2}\cosh2t$&$\sqrt{\frac{(n+\alpha)(n+\beta)(2n+\alpha+\beta-1)}{n(n+\alpha+\beta)(2n+\alpha+\beta+1)}}$&                          \\  \cline{2-2} \cline{5-5}
           &                               &                       &             & $-4(\alpha+\beta+n+m)(n-m+1)$                                                      \\
Natanzon   &$-\frac{1}{2}<x<+\frac{1}{2}$  &                       &             &                                                                                     \\
           &                               &                       &                                                            &                                      \\  \hline
\end{tabular}
\caption{ Some of known shape invariant potentials and their corresponding
 raising and lowering operators }
\end{table}

\begin{table}
\begin{tabular}{|c|c|c|c|}        \hline
$A(x)$ and    & $\tilde{A}$ and $\tilde{B}$       & $\omega_c(E)$    &$G$ and $a_{n}/a_0$\\
Name          &                                   &                  &\\  \hline
$1$           &                                   &                  &\\
              &$\frac{\alpha x}{2}+\frac{d}{dx}$  &                  &$x_0p_0$\\
              &                                   &$\omega_c=\alpha$ &\\\cline{2-2} \cline{4-4}
shifted       &                                   &                  &\\
oscillator    &$\frac{\alpha x}{2}-\frac{d}{ax}$  &                  &$\frac{(k_0/\alpha)^n}{\sqrt{\Gamma(n+1)}}$\\
              &                                   &                  &\\  \hline
$x$           & $\frac{\beta x}{2}+x\frac{d}{dx}$ &                  &$\frac{x_0 p_0(\alpha + m - \frac{1}{2})}{\beta}\times$\\
              & $-\frac{1}{2}(2n+\alpha-m+\frac{1}{2})$&             &$(1+\frac{2H}{(\alpha+m-1)\beta})$ \\
              &                                   &$\omega_c=\beta$  & \\  \cline{2-2} \cline{4-4}
three         & $\frac{\beta x}{2}-x\frac{d}{dx}$ &                  &\\
dimensional   & $-\frac{1}{2}(2n+\alpha-m+\frac{3}{2})$&             &$\frac{(-k_0)^n}{\sqrt{\Gamma(n+1) \Gamma(n+ \alpha +1)}}$ \\
oscillator    &                                   &                  &\\  \hline
$x^2$         &$-(n+\frac{\alpha+1}{2})x+x^2\frac{d}{dx}$&           &\\
              &$-\frac{\beta}{2}-\frac{\beta(m-n)}{2n+\alpha}$&$\omega_c=$&$x_0p_0e^{2t}$\\
              &                                   &$2\sqrt{\frac{\alpha^2}{4}+\frac{4m^2-1}{8}-E}$&\\  \cline{2-2} \cline{4-4}
Morse         &$-(n+\frac{\alpha-1}{2})x-x^2\frac{d}{dx}$ &           &\\
              &$+\frac{\beta}{2}-\frac{\beta(m+n+\alpha)}{2n+\alpha}$& &$(2k_0)^n \Gamma(n+\alpha/2+1)\sqrt{\frac{\Gamma(n+\alpha+1)}{\Gamma(n+1)}}$\\
              &                                   &                    &   \\ \hline
$1+x^2$       &$-(n+\alpha +\frac{1}{2})x+(1+x^2)\frac{d}{dx}$&$s_1=\alpha^2+\frac{4m^2-1}{8}$&    \\
              &$-\frac{\beta}{2}-\frac{\beta}{2}\frac{m-n}{n+\alpha}$&$s_2=(2m-1)(\alpha-\frac{1}{2})-E$  &$x_0p_0\cosh^2(t)$  \\
              &                                   &  & \\  \cline{2-2} \cline{4-4}
Scarf II      &$-(n+\alpha -\frac{1}{2})x-(1+x^2)\frac{d}{dx}$&$\omega_c=2\sqrt{s_1+s_2}$ &  \\
(hyperbolic)  &$+\frac{\beta}{2}-\frac{\beta}{2}\frac{m+n+2\alpha}{2(n+\alpha)}$&       &$(2k_0)^n\frac{\Gamma(n+2\alpha+1)}{\Gamma(n+\alpha+1)\Gamma(n+m+2\alpha)}$ \\
              &                                                            &            &    \\ \hline
$x(1-x)$      &$(n+ \frac{\alpha + \beta+1}{2})x+x(1-x) \frac{d}{dx}- \frac{2 \alpha+1}{4}$&$s_1=\frac{(\alpha+\beta)^2}{4}$&    \\
              &$-\frac{n(n+ \beta)}{2n+ \alpha + \beta}- \frac{m(\alpha - \beta)}{2(2n+ \alpha+ \beta)}$&$s_2=(m-\frac{1}{2})(\alpha+\beta-1)$&$x_0p_0\cos^2(t)$  \\
              &                                   &$s_3=-\frac{4m^2-1}{8}+E$&\\  \cline{2-2} \cline{4-4}
Scarf I &$(n+ \frac{\alpha + \beta-1}{2})x-x(1-x) \frac{d}{dx}+ \frac{2\alpha+1}{4}$&$\omega_c=2\sqrt{s_1+s_2+s_3}$&$(-2k_0)^n\Gamma(n+\frac{\alpha+\beta}{2})\sqrt{\frac{\Gamma(n+\frac{\alpha+\beta-1}{2})}{\Gamma(n+\frac{\alpha+\beta+1}{2})}}\times$\\
(trigonometric)&$-\frac{(n+ \alpha)(n+ \alpha + \beta)}{2n+ \alpha + \beta}-\frac{m(\alpha-\beta)}{2(2n+ \alpha+ \beta)}$&&\\
              &                                   &                       &$\sqrt{\frac{1}{\Gamma(n+1)\Gamma(n+\alpha+1)\Gamma(n+\beta+1)\Gamma(n+\alpha+\beta+1)}}$\\ \hline
$x^2-1$       &$-(n+\frac{\alpha+\beta+1}{2})x+(x^2-1)\frac{d}{dx}$&$s_1=\frac{(\alpha+\beta)^2}{4}$        & \\
              &$-\frac{\alpha-\beta}{2}-\frac{(\alpha-\beta)(m-n)}{2n+\alpha+\beta}$&$s_2=\frac{4m^2-1}{8}-E$&$x_0p_0\sinh^2(t)$  \\
              &                                   &$s_3=\frac{2m-1}{2}(\alpha+\beta-1)$&\\  \cline{2-2} \cline{4-4}
generalized &$-(n+\frac{\alpha+\beta-1}{2})x-(x^2-1)\frac{d}{dx}$&$\omega_c=2\sqrt{s_1+s_2+s_3}$&$\Gamma(n+\frac{\alpha+\beta}{2})\sqrt{\frac{\Gamma(n+\frac{\alpha+\beta-1}{2})}{\Gamma(n+\frac{\alpha+\beta+1}{2})}}\times$\\
P\"{o}schl-teller &$+\frac{\alpha-\beta}{2}-\frac{(\alpha-\beta)(m+n+\alpha+\beta)}{2n+\alpha+\beta}$&&$\sqrt{\frac{1}{\Gamma(n+1)\Gamma(n+\alpha+1)\Gamma(n+\beta+1)\Gamma(n+\alpha+\beta+1)}}$\\
              &                                   &                       &           \\  \hline
$1-x^2$       &$(n+\frac{\alpha + \beta}{2})x+(1-x^2)\frac{d}{dx}$ & $s_1=\frac{(\alpha+\beta)^2}{4}$& \\
              &$+\frac{\alpha - \beta}{2}-\frac{(n-m)(\alpha - \beta)}{2n+\alpha+\beta}$&$s_2=E-\frac{4m^2-1}{8}$&$x_0p_0\sin^2(t)$  \\
              &                                   &$s_3=\frac{2m-1}{2}(\alpha+\beta-1)$&\\  \cline{2-2} \cline{4-4}
         &$(n+\frac{\alpha + \beta}{2})x-(1-x^2)\frac{d}{dx}$&$\omega_c=2\sqrt{s_1+s_2+s_3}$&$(2k_0)^n\Gamma(n+\frac{\alpha+\beta}{2})\sqrt{\frac{\Gamma(n+\frac{\alpha+\beta-1}{2})}{\Gamma(n+\frac{\alpha+\beta+1}{2})}}\times$\\
              &$+\frac{\alpha - \beta}{2}+\frac{(n+m+\alpha+\beta)(\alpha - \beta)}{2n+\alpha+\beta}$&&$\sqrt{\frac{1}{\Gamma(n+1)\Gamma(n+\alpha+1)\Gamma(n+\beta+1)\Gamma(n+\alpha+\beta+1)}}$\\
     &                               &                       &                                                            \\   \hline
\end{tabular}
\end{table}

\begin{table}
\begin{tabular}{|c|c|c|c|}        \hline
$A(x)$ and    & $\tilde{A}$ and $\tilde{B}$       & $\omega_c(E)$    &$G$ and $a_{n}/a_0$\\
Name          &                                   &                  &\\  \hline

$4x^2-1$      &$-4(n+\frac{\alpha+\beta+1}{2})x+(4x^2-1)\frac{d}{dx}$&$s_1=\alpha+\beta)^2$&                                        \\
        &$-(\alpha-\beta)-\frac{2(\alpha-\beta)(m-n)}{2n+\alpha+\beta}$&$s_2=\frac{4m^2-1}{8}-E$&$x_0p_0\sinh^2(2t)$  \\
        &                               &$s_3=2(2m-1)(\alpha+\beta-1)$&\\  \cline{2-2} \cline{4-4}
     &$-4(n+\frac{\alpha+\beta-1}{2})x-(4x^2-1)\frac{d}{dx}$&$\omega_c=4\sqrt{s_1+s_2+s_3}$&$(k_0/4)^n\Gamma(n+\frac{\alpha+\beta}{2})\sqrt{\frac{\Gamma(n+\frac{\alpha+\beta-1}{2})}{\Gamma(n+\frac{\alpha+\beta+1}{2})}}\times$\\
Natanzon  &$+(\alpha-\beta)-\frac{2(\alpha-\beta)(m+n+\alpha+\beta)}{2n+\alpha+\beta}$& &$\sqrt{\frac{1}{\Gamma(n+1)\Gamma(n+\alpha+1)\Gamma(n+\beta+1)\Gamma(n+\alpha+\beta+1)}}$\\
              &                               &                       &                                                             \\  \hline
\end{tabular}
\end{table}

\section{The Most General Minimum Uncertainty States}

The coherent and squeezed states are generally the minimum uncertainty
states of general harmonic phase variables $\widehat{X}$ and $\widehat{P}$,
which can be obtained by solving the eigenfunction equation of the operators
$\widehat{X} + i \frac{<G>}{2(\Delta p)^2} \widehat{P}$
\cite{Ya:Niet,Do:Niet,Do:Niea,Do:Nieb}, that is, we solve

\setcounter{equation}{0}

\begin{equation}
(\widehat{X} + i \frac{<G>}{2(\delta p)^2} \widehat{P})\psi_{MUCS} = C\psi_{MUCS}
\end{equation}
where $G$ is proportional to the commutation of $\widehat{X}$ and $\widehat{P}$ as
$[\widehat{X},\widehat{P}]=iG$, and where $C=<\widehat{X}> + i \frac{<G>}{2(\Delta p)^2} <\widehat{P}>$.

In order to solve the Eq.$ (5-1)$, we expand $\psi_{MUCS}$ in terms of $\psi_n^m$,
 the eigenstate of Schrodinger equation $(2-4)$,

\begin{equation}
\psi_{MUCS} =\sum_{j=m}^{\infty} a_j \psi_j^m(x).
\end{equation}
Inserting the above expansion in Eq.$(5-1)$  and using the independence of the eigenstates $\psi_j^m$,
we get a recursion relation
between the coefficients $a_{j}$.
This is possible if we know how the adjoint operators
$\tilde{A}^{\dagger}$ and $\tilde{B}^{\dagger}$ act over $\psi_j^m$
by writing them in terms of the raising and lowering operators
$\tilde{A}$ and $\tilde{B}$ as

\[\tilde{A}^{\dagger}=(\frac{f_1-f_3-A^{\prime\prime}(x)}{f_1+f_3})\tilde{A}+(\frac{2f_1-A^{\prime\prime}(x)}{f_1+f_3})\tilde{B}+ \]
\[\frac{(f_2-A^{\prime}(0))(f_1+f_3)-(f_1-A^{\prime\prime}(x))(f_2+f_4)+\frac{1}{2}(f_2-f_4)(f_1+f_3)-\frac{1}{2}(f_1-f_3)(f_2+f_4)}{f_1+f_3}  \]

\[\tilde{B}^{\dagger}=(\frac{2f_3+A^{\prime\prime}(x)}{f_1+f_3})\tilde{A}+(\frac{f_3-f_1+A^{\prime\prime}(x)}{f_1+f_3})\tilde{B}+ \]
\[\frac{(f_4+A^{\prime}(0))(f_1+f_3)-(f_3+A^{\prime\prime}(x))(f_2+f_4)+\frac{1}{2}(f_1-f_3)(f_2+f_4)-\frac{1}{2}(f_2-f_4)(f_1+f_3)}{f_1+f_3}  \]
Therefore, the most general quantum harmonic phase coordinate $\widehat{X}$ and
momentum $\widehat{P}$, given in Eq.$(5-1)$, can be expressed in terms of the raising
and lowering operators.

\[\widehat{X} =2x_0(\tilde{A} + \tilde{B}) \]

\[\frac{2i}{p_0}\widehat{P} =\frac{4f_3+2A^{\prime\prime}(x)}{f_1+f_3}\tilde{A}+\frac{-4f_1+2A^{\prime\prime}(x)}{f_1+f_3}\tilde{B}+ \]
\[\frac{2(f_1+f_3)(f_4-f_2)+2(f_2+f_4)(f_1-f_3)+2A^{\prime}(0)(f_1+f_3)-2A^{\prime\prime}(x)(f_2+f_4)}{f_1+f_3}\]

Substituting the operators $\widehat{X}$ and $\widehat{P}$, written only in terms of
the raising and lowering operators as above, in Eq.$(5-1)$ and using the
expansion of $\psi_{MUCS}(x)$ in terms of $\psi_n^m(x)$, we obtain the following
recursion relation for $a_{n+m}$:

\[a_{n+m+1}=\frac{1}{2x_0+\frac{p_0<G>}{4(\Delta p)^2}\frac{4f_3+2A^{\prime\prime}(x)}{f_1+f_3}}\frac{\mu_{n+1,m}}{E(n+1,m)}\times \]
\begin{equation}
[(C-\frac{p_0<G>}{4(\Delta p)^2}g)a_{n+m} - (2x_0-\frac{p_0<G>}{4(\Delta p)^2}\frac{4f_1-2A^{\prime\prime}(x)}{f_1+f_3})\mu_{n,m}a_{n+m-1}].
\end{equation}
where
\[g=\frac{2(f_1+f_3)(f_4-f_2)+2(f_2+f_4)(f_1-f_3)+2A^{\prime}(0)(f_1+f_3)-2A^{\prime\prime}(x)(f_2+f_4)}{f_1+f_3}.\]

In principle, by iterating the recursion relation (5-3) we can determine $a_{n+m}$
in terms of $a_0$ and $a_{m+1}$. In general, due to the appearance of $a_{n+m}$ and
$a_{n+m-1}$ on the right hand side of this recursion relation, we are not able
to obtain a closed form for the coefficient $a_{n+m}$ in terms of $a_m$ and $a_{m+1}$.
Note that, by an appropriate choice of the parameters appearing in MUCS,
as e.g. taking
such as the averge of the quantum phase coordinates,
 the average of the commutator
of the quatum phase coordinates or the squeezing percentage that is the ratio standard deviations,
$(2x_0-\frac{p_0<G>}{4(\Delta p)^2}\frac{4f_1-2A^{\prime\prime}(x)}{f_1+f_3})$
vanishes.
We iterate
\[\frac{a_{n+m+1}}{a_{n+m}}=\frac{(C-\frac{p_0<G>}{4(\Delta p)^2}g)}{2x_0+\frac{p_0<G>}{4(\Delta p)^2}\frac{4f_3+2A^{\prime\prime}(x)}{f_1+f_3}}\frac{\mu_{n+1,m}}{E(n+1,m)}\]
to obtain $a_{n+m}$ only in terms of $a_m$:
\begin{equation}
\frac{a_{n+m}}{a_m}=k_0^n\prod_{j=m}^{n+m}\frac{\mu_j}{E(j)}
\end{equation}
with
\[k_0=\frac{(C-\frac{p_0<G>}{4(\Delta p)^2}g)}{2x_0+\frac{p_0<G>}{4(\Delta p)^2}\frac{4f_3+2A^{\prime\prime}(x)}{f_1+f_3}}.\]
Substituting the coefficient $a_{n+m}$, given in Eq.$(5-4)$, into the expansion $\psi_{MUCS}$, we get

\begin{equation}
\psi_{MUCS}=\sum_{n=m}^{\infty}a_mk_0^n\prod_{j=m}^{n+m}\frac{\mu_j}{E(j)}\psi_n^m.
\end{equation}

In the rest of this section we investigate some important special cases:

a- For $A(x)=1$ we get the harmonic oscillator with eigenfunction
\[\psi_n(x)=e^{-\frac{\alpha x^2}{4}}H_n(x)\]
and the coherent states yield
\[\psi_{MUCS}(x)=e^{tx-t^2/2}\],
where $t=\frac{k_0}{\alpha}$.

b- $A(x)=x, \beta=1, m=0, and \alpha=\lambda+\frac{1}{2}$ lead to
\[\psi_n(x)=x^{\frac{\alpha+1/2}{2}}e^{-x/2}n!L_n^{\alpha}(x)\]
and the coherent states yield
\[\psi_{MUCS}(x)=x^{\frac{\alpha+1/2}{2}}e^{-x/2}I_{\lambda+1/2}(2\sqrt{-k_0x}),\]
in agreement with reference\cite{Do:Niea}.

c- $A(x)=x(1-x), \alpha=\beta=0, m=1/2-\lambda, and n=n+\lambda-1/2 $ lead to
\[\psi_n(x)=\sqrt{\cos(t)}P_{n+\lambda-1/2}^{1/2-\lambda}(\sin(t))\]
and the coherent states yield
\[\psi_{MUCS}(x)=\]
in agreement with reference\cite{Do:Niea}.

d- $A(x)=1+x^2, \alpha=-1/2, and \beta=0,  $ lead to
\[\psi_n(x)=P_{m-1}^{n}(\tanh(t))\]
and the coherent states yield
\[\psi_{MUCS}(x)=\]
in agreement with reference\cite{Do:Nieb}.

e- $A(x)=x^2, \beta=1, \lambda=n-\alpha/2$ lead
\[\psi_n(x)=x^{-\frac{\alpha+1}{2}}e^{-x/2}L_n^{-\alpha-1}(x)\]
and the coherent states yield
\[\psi_{MUCS}(x)=\]
in agreement with reference\cite{Do:Nieb}.
If the coeffient of $a_n$ in the recursion relation (5-3) vanishes, that is for:

\[ C=\frac{p_0 <G>}{4(\Delta p)^2 }g \]
we encounter another interesting special case. This is again possible by an
appropriate choice of the parameters of MUCS. In this case the MUCS can consist only of the odd or even associative
orthogonal polynomials. In particular, if we choose even master function together
with the corresponding even weight function, then the associative orthogonal
function will be even for even $n+m$ and will be odd for odd $n+m$, provided
that we choose a symmetric interval,that is $[-a,a]$. Therefore, MUCS is either
even or odd known as even or odd coherent (squeezed) states in the
special case of the harmonic Hamiltonian (Schrodinger-Cat states)
\cite{Al:Ros1,Al:Ros2,Al:Ros3,Al:Ros4}.
Finally we can choose the parameters of MUCS such that the coefficients of
$a_{n+m}$ and $a_{n+m-1}$ on the right hand side of the recursion (5-3) become null.
This means that in the expansion of MUCS, given in (5-2), only the ground state
of the Hamiltonian will remain, that is, the ground state of the system belongs
to its MUCS ensemble.
At the end of this section it should be remined that,
eventhough the closed form of MUCS is only available in seldom special
case.  But this is not a serious prblem, since one can iterate the recursion
relation (5-3) numerically, to calculate the MUCS and related average quantities
over these states which is under separate investigation.

\begin{center}
\section{Eigenstates of Annihilation Operators}
\end{center}

It is well-known that the coherent states of harmonic oscillators are eigenstates
of the annihilation operator $a$, that is, we have
\[a \mid \alpha >=\alpha \mid \alpha >.\]
Here we try to generalize this idea to the general shape invariant potentials ($V_m(x(\xi))$).
This is possible if we can find the eigenstate of annihilation operator
($A \mid \alpha >=\alpha \mid \alpha >$) associated with these potentials.
Here in this section, by multiplying the annihilation operator by an appropriate
function of Hamiltonian and using the result of reference \cite{Lo:Jafy},
we get a closed form for their eigenstates.
Denoting  the eigenstates of the annihilation operator $A$ by
$\psi_{AOCS}(\beta_0,x)$, we have the following
eigen-equation

\setcounter{equation}{0}

\[F(g^{-1}(H))A \psi_{AOCS}(\beta_0,x)=\beta_0 \psi_{AOCS}(\beta_0,x),   \]
where $F(g^{-1}(H))$ is an arbitrary function of Hamiltonian which
is to be determined below.
Expanding $\psi_{AOCS}(\beta_0,x)$ in terms of the associated orthogonal
functions $\phi_{n,m}(x)$ and using the relation

\[F(g^{-1}(H))A \sum_{n=0}^{\infty}C_n \phi_{n,m}=\beta_0 \sum_{n=0}^{\infty}C_n \phi_{n,m},\]
we get the following recursion relation

\begin{equation}
C_{n+1}\frac{F(n+1)E(n+1,m)}{\mu_{n+1,m}}=\beta_0 C_n.
\end{equation}
Substituting for $\mu_{n+1,m}$ we get
\[\frac{C_{n+1}}{C_n}=\beta_0 \frac{\frac{-1}{2}nA^{\prime\prime}(x) -B_0}{E(n+1,m)F(n+1)} \frac{a_n}{a_{n+1}}\]
with
\[B_0=\frac{1}{2}(n+1)A^{\prime\prime}(x)+(\frac{A(x)W^{\prime}(x)}{W(x)})^{\prime}.\]
By defining $F(g^{-1}(H))$ as
\[F(n+1)=\frac{(n+1)(\frac{-1}{2}nA^{\prime\prime}(x) -B_0)}{E(n+1,m)}\]
we can solve the recursive relation $(6-1)$ if we
choose $C_n=\frac{\lambda^n}{n!a_n}$ and $\lambda=\beta_0$.
Substituting these in the expansion of $\psi_{AOCS}(\beta_0,x)$ we get

\[\psi_{AOCS}(\beta_0,x)=\sum_{n=0}^{\infty}\frac{\beta_0^n}{n!a_n}\phi_{n,m}.\]
Now, using the result of reference \cite{Lo:Jafy} we obtain
\[\psi_{AOCS}(\beta_0,x)=(-1)^m(A(x))^{\frac{m}{2}} (\frac{d}{dx})^m (\frac{W(z)}{W(x)} \frac{dz}{dx})\]
with $z=x+tA(z)$.
From the result thus obtained it is clear that in general, the $\psi_{AOCS}$
states are different from the corresponding $\psi_{MUCS}$ states and they only coincide
in the case of harmonioc oscillator. To see their difference more
explicitly, using the result of reference\cite{Lo:Jafy}, we give
in the rest of this section some of the $\psi_{AOCS}$
states.

I-The choice of $A(x)=x, \beta=1, m=0,\alpha=\lambda+\frac{1}{2}$ leeds
\[\psi_n(x)=x^{\frac{\alpha+1/2}{2}}e^{-x/2}n!L_n^{\alpha}(x)\]
and the using coherent states
\[\psi_{MUCS}(x)=x^{\frac{\alpha+1/2}{2}}e^{-x/2}I_{\lambda+1/2}(2\sqrt{-k_0x})\]
which is in agreement with reference\cite{Do:Niea}.

II-The choice of $A(x)=x(1-x), \alpha=\beta=0, m=1/2-\lambda, n=n+\lambda-1/2 $ leeds
\[\psi_n(x)=\sqrt{\cos(t)}P_{n+\lambda-1/2}^{1/2-\lambda}(\sin(t))\]
and the using coherent states
\[\psi_{MUCS}(x)=\]
which is in agreement with reference\cite{Do:Niea}.

III-The choice of $A(x)=1+x^2, \alpha=-1/2, \beta=0,  $ leeds
\[\psi_n(x)=P_{m-1}^{n}(\tanh(t))\]
and the using coherent states
\[\psi_{MUCS}(x)=\]
which is in agreement with reference\cite{Do:Nieb}.

IV-The choice of $A(x)=x^2, \beta=1, \lambda=n-\alpha/2$ leeds
\[\psi_n(x)=x^{-\frac{\alpha+1}{2}}e^{-x/2}L_n^{-\alpha-1}(x)\]
and the using coherent states
\[\psi_{MUCS}(x)=\]
which is in agreement with reference\cite{Do:Nieb}.

\section{Time Evolution of the Minimum Uncertainty States}

In this section we investigate the time evolution of the generalized
Harmonic quantum
phase variable $\widehat{X}$ and $\widehat{P}$. Since they do not have any explicit time dependence,
thus their time dependence can be written as

\setcounter{equation}{0}

\renewcommand{\theequation}{\thesection-\arabic{equation}{a}}
\begin{equation}
\widehat{X}(t)=e^{\frac{iHt}{\hbar}} \widehat{X} e^{-\frac{iHt}{\hbar}},
\end{equation}
\setcounter{equation}{0}
\renewcommand{\theequation}{\thesection-\arabic{equation}{b}}
\begin{equation}
\widehat{P}(t)=e^{\frac{iHt}{\hbar}} \widehat{P} e^{-\frac{iHt}{\hbar}},
\end{equation}

\renewcommand{\theequation}{\thesection-\arabic{equation}}

\setcounter{equation}{1}

which follows from the Heisenberg equations of motion

\[\dot{\widehat{X}}=\frac{1}{i\hbar}[\widehat{X},H]=\frac{\widehat{P}}{m}\]

\[\dot{\widehat{P}}=\frac{1}{i\hbar}[\widehat{P},H]=\widehat{X}B_1(H)+i\widehat{P}B_0\]

with $B_1(H)$  and $B_0$ defined as:

\[B_0=-\frac{1}{2m}\hbar A^{\prime\prime}(x)\]

$$
B_1=A^{\prime\prime}(H-\gamma +\frac{2m-1}{4}A^{\prime\prime})
+\frac{1-2m}{2}A^{\prime\prime}(\frac{AW^{\prime}}{W})^{\prime}
-\frac{1}{2}((\frac{AW^{\prime}}{W})^{\prime})^2
-\frac{4m^2-1}{8}A^{\prime\prime}.
$$

Now, using the Baker-Hausdorf formula in  Eq.$(7-1a)$, we can calulate $\widehat X(t)$ :

\[\widehat{X}(t)=\sum_{n=0}^{\infty}(\frac{it}{\hbar})^n\frac{1}{n!}X_n\]
with
\begin{equation}
X_n=\widehat{X}f_n(H)+\widehat{P}g_n(H)
\end{equation}

By iterating the following recursion relations

\begin{equation}
f_{n+1}(H)=\frac{\hbar}{i}B_1(H)g_n(H)
\end{equation}

\begin{equation}
g_{n+1}(H)=\frac{\hbar}{i}(\frac{1}{m}f_n(H)+(iB_0)g_n(H))
\end{equation}
with

we get the following expression for the function $g_n(H)$

\[g_n(H)=Ar_+^n+Br_-^n\]
where $r_- , r_+, A$ and $B$ are

\[r_+=\frac{1}{2}\hbar B_0 + \frac{1}{2}\hbar \sqrt{B_0^2-4B_1(H)/m}=\hbar\omega_0+\hbar\omega_H\]
\[r_-=\frac{1}{2}\hbar B_0 - \frac{1}{2}\hbar \sqrt{B_0^2-4B_1(H)/m}=\hbar\omega_0-\hbar\omega_H\]
\[A=-B=\frac{-i\hbar}{2m\omega_H}\]
with $\omega_0 $ and $\omega_H $ defined as
$$
\omega_0=B_0/2 \;\;\;\;\ {\em and} \;\;\;\; \omega_H=\sqrt{B_0^2-4B_1(H)/m}/2,
$$
respectively.
Substituting the result thus obtained for $g_n(H)$ in (7-4a) we can determine
$f_n(H)$.\\
Now, inserting all these in (7-3), we obtain the closed form for
$\widehat{X}(t)$ and $\widehat{P}(t)$ as follows:

\[\widehat{X}(t)=\widehat{X}e^{i\omega_0 t}[\cos(\omega_H t)-i\frac{\omega_0}{\omega_H}\sin(\omega_H t)]\]

\begin{equation}
+\widehat{P}e^{i\omega_0 t}2\frac{\omega_0}{\omega_H}\sin(\omega_H t)
\end{equation}

Starting from the formula (7-1b) and performing  some calculation  which is similar
to the above calculation, we will obtain

\[\widehat{P}(t)=\widehat{P}e^{i\omega_0 t}[\cos(\omega_H t)+i\frac{\omega_0}{\omega_H}\sin(\omega_H t)]\]
\begin{equation}
+\widehat{X}e^{i\omega_0 t}(2\frac{\omega_0}{\omega_H})\sin(\omega_H t)
\end{equation}

Again the result thus obatained are in agreement with refernces\cite{Do:Niea,Do:Nieb}
for the special caes provided that we insert the corresponding eigen value of
the operators of eigen frequency $\omega_H $, given in Table I.
The time dependence of the generlized quantum phase coordinates given in (7-5) and
(7-6) is almost similar to the time evolution of the phase coordinates of the
qunantum oscillator except for the appearance of the ${\omega_0}$ and also
the energy depenence of the $\omega_H$. We see that in case of the general
shape invarint hamiltonian the frequency is a hamiltonian dependent operator
and it is only constant in special case of oscillator.

\begin{center}
\section{CONCLUSION}
\end{center}

In this paper a general algorithm has been given for the generation of the
minimum uncertainty coherent and squeezed states in some one-dimensional
hamiltonians with shape invariant  potential, obtained from the master
function. It looks like that the shape invariance symmetry
of these hamiltonian might be the reason for the observation the MUCS.
Since Solvability of these quantum systems are mainly due to the
existance of this symmetry \cite{Lo:Jafa,La:AJaa}. But this not the only reason
, Actually the main role belongs the existence of the lowering and raising
operators or ladder ones \cite{Ya:Niet,Do:Niet,Do:Niea,Do:Nieb},
which map different energy eigenstates
of a given hamiltonian into each other. As quoted in the introduction,
the coherent and squeezed states generated by harmonic oscillator have
already play such an importantat role in different branches of physics.
Definetly the MUCS have been generated in refrence \cite{Do:Niet} and here will
soon play very important role in almost all branches of physics. Therefore,
it deserve to find all other hamiltonian which can generate MUCS, which is
under investigation.

\begin{center}
{\bf {\large  ACKNOWLEDGEMENT}}
\end{center}

We wish to thank  Dr. S. K. A. Seyed Yagoobi for his careful reading the
article and for his constructive comments.

\end{document}